\documentclass{aa} 
\usepackage{graphics}
\usepackage{natbib}
\bibpunct{(}{)}{;}{a}{}{,}
\usepackage{longtable}
\usepackage{verbatim}
\usepackage{txfonts}
\begin{document}
\title{Identification of Emission Lines in the Low-Ionization Strontium Filament
Near Eta Carinae\thanks{Based on observations made with the
NASA/ESA {\it Hubble Space Telescope}, obtained at the Space
Telescope Science Institute, which is operated by the Association
of Universities for Research in Astronomy, Inc., under NASA
contract NAS5-26555. }}
\author{H. Hartman \inst{1} \and T. Gull \inst{2} \and S. Johansson \inst
{1} \and N. Smith \inst{3}\thanks{Hubble fellow} \and HST Eta~Carinae Treasury Project Team
\thanks{This research is partly based on data from the Eta Carinae Hubble Space Telescope Treasury 
project via grant no.\ GO-9420 from Space Telescope Science Institute.}
}
\offprints{Henrik Hartman, \email{henrik.hartman@astro.lu.se}}
\institute{Atomic Astrophysics, Lund Observatory, Lund University,
Box 43, SE-221\,00 Lund, Sweden \and Laboratory for Astronomy and
Solar Physics, Code 681, Goddard Space Flight Center, Greenbelt,
MD, USA, 20771 \and CASA, 389 UCB, University of Colorado, Boulder, CO, 80309 }
\date{Received <date> / Accepted <date>}
\titlerunning{Emission Lines in the Strontium Filament}
\authorrunning{Hartman et al.}

\abstract{We have obtained deep spectra from 1640 to 10100\AA\
with the Space Telescope Imaging Spectrograph (STIS) of the
Strontium Filament, a largely neutral emission nebulosity lying
close to the very luminous star Eta Carinae and showing an uncommon 
spectrum. Over 600 emission
lines, both permitted and forbidden, have been identified. The
majority originates from neutral or singly-ionized iron group
elements (Sc, Ti, V, Cr, Mn, Fe, Co, Ni). Sr is the only neutron
capture element detected. The presence of Sr II, numerous strong
\ion{Ti}{ii} and \ion{V}{ii} lines and the dominance of \ion{Fe}{i} over \ion{Fe}{ii} are
notable discoveries. While emission lines of hydrogen, helium, and
nitrogen are associable with other spatial structures at other
velocities within the Homunculus, no emission lines from these
elements correspond to the spatial structure or velocity of the
\ion{Sr}~Filament. Moreover, no identified \ion{Sr}~Filament
emission line requires an ionization or excitation energy above
approximately 8 eV. Ionized gas extends spatially along the
aperture, oriented along the polar axis of the Homunculus, and in
velocity around the Strontium Filament. We suggest that the
Strontium Filament is shielded from ultraviolet radiation at
energies above ~8 eV, but is intensely irradiated by the central
star at wavelengths longward of 1500\AA. }

\maketitle 
\begin{figure*}
\resizebox{!}{8cm}{\includegraphics{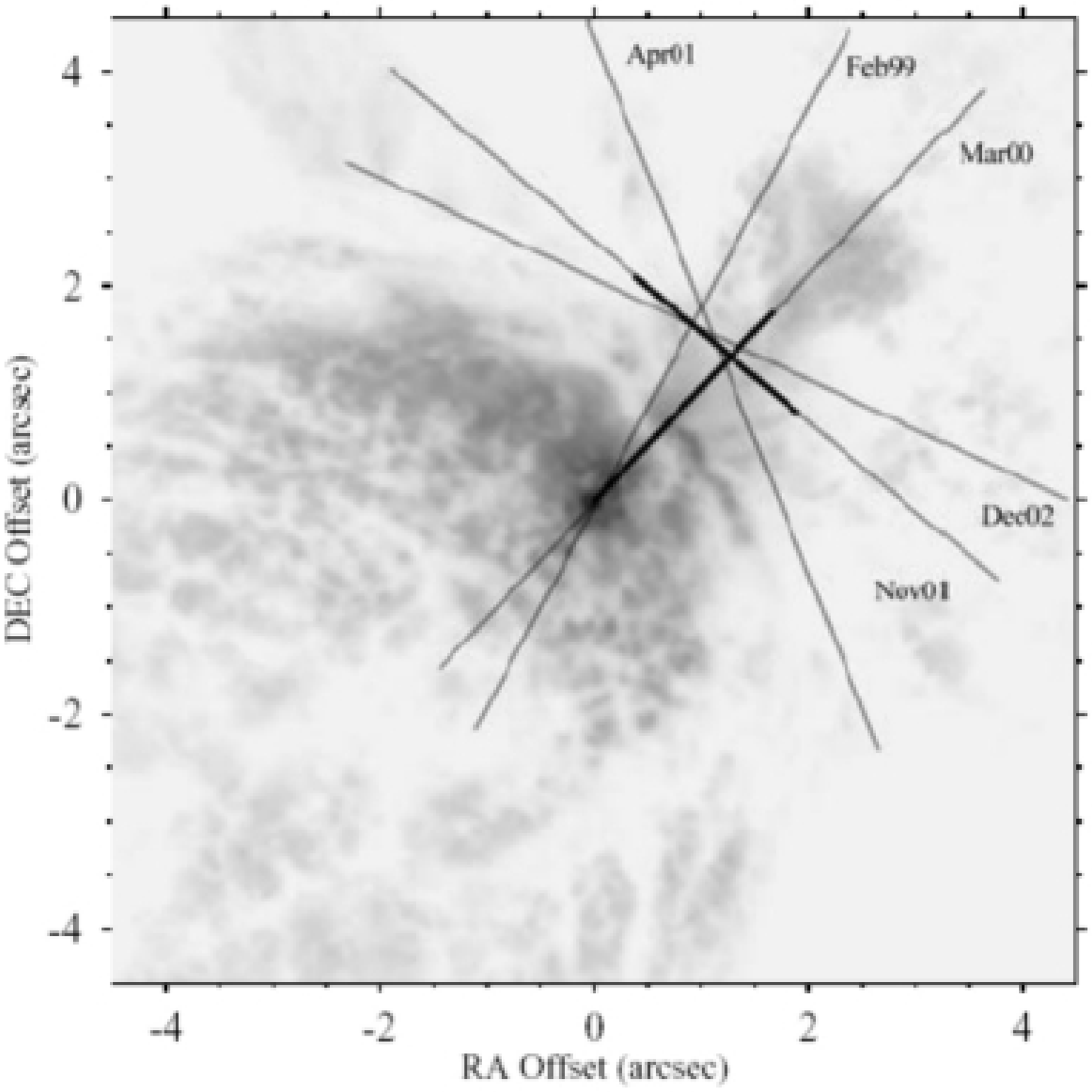}} 
\resizebox{!}{8cm}{\includegraphics{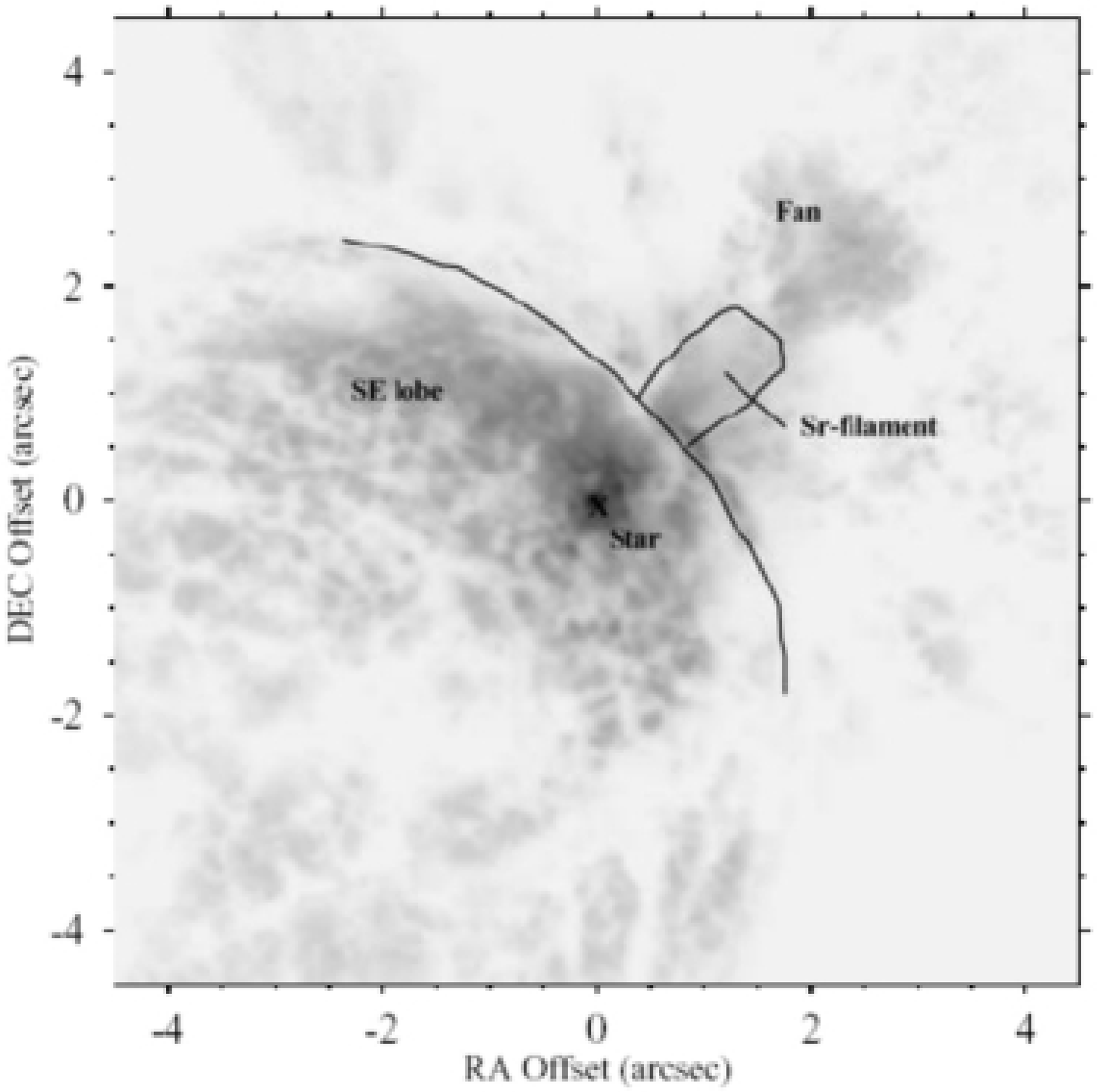}} 
\caption{ACS/HRC image of $\eta$ Car and the Homunculus \citep{SMG04} with
the aperture positions superimposed (See Table 1). The sections plotted in Fig. 2 and 3 
are marked with bold lines. The March 2000
aperture represents the aperture centered on $\eta$ Car. At this same epoch, multiple
exposures were recorded with offsets orthogonal to the aperture to
map the nebular structure \citep{IGD03}. We have used
these same spectra to estimate the size of the \ion{Sr}~Filament
in \ion{Sr}{ii}, sketched in the right panel. \label{picslits}}
\end{figure*}
\section{Introduction}
The Luminous Blue Variable star (LBV) Eta Carinae ($\eta$ Car)
experienced a major outburst in the 1840's with a secondary
outburst in the 1890's \citep[][ and references therein]{DH97}.
Several solar masses of material were
ejected, which is now directly seen as the expanding Homunculus
\citep{MDB98,SGH03}
and the
Little Homunculus \citep{IGD03}.
Although the central
star provides 5x10$^6$ L$_{\odot}$ at a characteristic temperature
of $25,000\degr $K, most of the gas in the Homunculus is neutral
\citep{DSG01}. The hollow bipolar lobes with an
intervening equatorial skirt are seen primarily by dust-scattered
radiation from $\eta$ Car. STIS CCD spectra, recorded with a long
aperture ($52\arcsec \times 0\farcs2$), revealed a thin, interior
skin in the light of [\ion{Fe}{ii}] and [\ion{Ni}{ii}], but no
\ion{H}{ii} nor \ion{He}{i}, emission.  
\citet{DSG01} used
these lines, and the absorption H and K lines of \ion{Ca}{ii}, to
trace the inner, neutral surface of the Homunculus. Ground-based
observations in the near-infrared \citep{S02}
revealed molecular hydrogen in the cool exterior
of the Homunculus shell. Lines of \ion{Fe}{ii}, H$\alpha$ and [\ion{Ni}{ii}]
revealed an internal emission nebulosity, called the Little
Homunculus 
\citep{GI01,IGD03}.
Several bright, ionized emission structures exist very close to
$\eta$ Car, known as the Weigelt Blobs B, C and D (Weigelt \&
Ebersberger 1986), are located within $0\farcs1$ to $0\farcs3$
from the central star.

\citet{D96} identified the
5.52-year period in the high-excitation nebular and stellar
emission lines of $\eta$ Car and its surrounding nebulosity. In a
long-term monitoring series of programs to understand this
variability, the Weigelt blobs B and D, along with $\eta$ Car,
have been observed with HST/STIS at nearly annual intervals since
1998 \citep{DIG99,GID01}.
Over 2000 emission lines of
the Weigelt blobs B and D were identified in the spectrum between
1700\AA\ and 10300\AA\ by \citet{ZPHD01}. Changes between the
spectroscopic minimum in 1998 and the broad maximum during 1999
and 2000 demonstrate that lines of higher ionization disappear
during the spectroscopic minimum only to reappear as the system
recovers. Many \ion{H} Ly$\alpha$-pumped \ion{Fe}{ii} lines
appear during the maximum 
\citep{JH93,ZPHD01}, and disappear during the minimum. The
\ion{Fe}{ii} 2507, 2509 \AA\ lines are the most enhanced of these
fluorescence lines, and they feed long-lived Fe II states involved
in a closed radiative cycle showing stimulated emission \citep{JL03}.
\citet{VGB02} used the CLOUDY model to
predict the optical Fe II emission fluxes of the Weigelt B and D
blobs during the spectroscopic minimum event of 1998.

During a preliminary test for the Homunculus mapping program in
the 6400-7000 \AA\ region planned as a STIS GTO Key Project (HST
proposal 8483), we noticed some very faint, narrow emission lines
located 1.5\arcsec\ to the Northwest of $\eta $ Car. The 1\arcsec\ long
emission filament appeared not to be associated with any known
structure in $\eta$ Car. Yet the spatial and velocity structure
was similar for these lines and
indicated that they must originate from the same volume.
\citet{ZGH01} identified twenty of twenty one lines in the
6400-7000A\ region, all originating from a structure moving at
-100 km/sec. As the most spectacular discovery was the first
identification of two [\ion{Sr}{ii}] lines, the filament became
known as the \ion{Sr}~Filament. Peculiarly, no lines of hydrogen
or helium were identified in the spectrum of this system. While
lines of \ion{Fe}{ii} were not identified, lines of \ion{Fe}{i}
were. However, from the identifications of this wavelength limited
spectrum it was not possible to conclude whether these emission
lines were due to a selective excitation mechanism or to different
elemental abundances.

The limited spectrum of the \ion{Sr}~Filament differed remarkably
from similar spectra of other emission line nebular structures
around $\eta$ Car. This fact led to additional observations and
line identifications in other wavelength regions. In the present
paper we report on all HST observations obtained so far of the
\ion{Sr}~Filament and tabulate all the measured emission lines.
Nearly 600 lines have been identified, and only a few strong lines remain
unidentified. We also discuss the peculiarities found in the
spectrum in terms of apparent enhancements and depletions in
elemental abundances, as well as clear indications of special
ionization and excitation conditions in the filament.

\section{Observations}
The initial 6480 to 7000\AA\ spectrum of the \ion{Sr}~Filament
differed markedly from spectra of other emission line structures
around $\eta$ Car, and indeed from spectra of other nebulae.
[\ion{Sr}{ii}] emission is not known to have been observed in
other emission nebulae. Given the uniqueness of this nebular
spectrum, we followed up with a series of observations, first to
detect \ion{Sr}{ii} lines near 4000\AA, then other nebular
emission lines, within visits scheduled for $\eta$ Car.
Information on these visits are listed in Table I. The two
resonance lines of \ion{Sr}{ii} at 4078 \AA\ and 4216 \AA\ were
observed in emission. \citet{BGI02} found the \ion{Sr}{ii} line
ratios to be consistent with a gas having electron densities of
$10^7$ cm$^{-3}$ in a predominantly neutral region. We extended
the spectral coverage across the entire range of the STIS CCD
(1640 to 10100 \AA), and examined the spectroscopic maps of the
Homunculus in spectral intervals containing H$\alpha$ and H$\beta$
to determine the spatial extent of the peculiar emission. The
observations were done during several HST visits. As the HST
spacecraft orientation changes throughout the year, we had to
accept observations through the long aperture at very different
position angles (Figure 1). When possible, the aperture was
centered on a common position offset $1\farcs5$ at Position Angle
$315\degr$ from $\eta$ Car. Enough overlap in repeat spectral
coverage (Table 1) allowed us to gain significant information on
the spatial extent of the \ion{Sr}~Filament.

\begin{table*}
\caption{Log of Observations with the STIS \label{tbl-1}}
\begin{tabular}{llllcl} \hline \hline
Date & HST& Offset & Pos Angle & Aperture & Spectral Coverage \\
& Proposal & from $\eta $ Car & Degrees & & \AA ngstroms  \\
& & R(\arcsec), $\theta$&(N through E)  \\ \hline
Feb 21, 1999 & 8036 & 0.4, 45$\degr$& -27.97$\degr$ &$52 \arcsec \times0\farcs1$ &6480-7000\AA $^1$\\
Mar 13, 2000 & 8327 & 0,0 $^2$& -41.14$\degr$ &$52 \arcsec \times 0\farcs2$F2&2480-2910\AA \\
& & & & & 3795-4335\AA \\
& & & & &4818-5100\AA \\
& & & & &6480-7565\AA $^{3,4}$\\
& & & & &9300-9600\AA \\
Mar 21, 2000 & 8483 & $N\times 0.1$, 55$\degr$$^5$ & -35.02$\degr$ & $52\arcsec\times 0\farcs 1$ & 4818-5100\AA \\
& & $N\times 0.25$, 55$\degr$$^5$ &  &  & 6480-7050\AA\\
Apr 17, 2001 & 8619 & $1\farcs 5$, 315$\degr$ & 22.06$\degr$ & $52\arcsec \times 0\farcs 2$& 4052-4593\AA $^6$\\
& & & & & 4818-5100\AA \\
& & & & & 6480-7050\AA $^3$\\
Nov 27, 2001 & 8619 & $1\farcs 5$, 315$\degr$ & -130.97$\degr$ & $52\arcsec\times 0\farcs 2$ & 2480-2633\AA \\
& & & & & 3022-10135\AA $^{3,4,6}$\\
Dec 16, 2002 & 9420 & $1\farcs 5$, 315$\degr$ & -114.94$\degr$ & $52\arcsec\times 0\farcs 2$ &1640-2492\AA \\
& & & & & 2897-3052\AA \\ \hline
\end{tabular}

$^1$ Initial discovery spectrum \citep{ZGH01}. 

$^2$ Mar 13, 2000 observations were accomplished with the F2 ($0\farcs 85
$) fiducial blocking the $\eta $ Car at the STIS entrance.

$^3$ Repeated spectra from 6480 - 7000\AA\ to check for variability.

$^4$ Repeated spectra at 6995-7565\AA\ to check variability

$^5$ Mapping spectra were recorded in the 4818-5100\AA\ with
$0\farcs1 $ offsets and 6480-7050\AA\ spectral range  with
$0\farcs25 $ offsets to map the spatial structure of the
Homunculus and the Little Homunculus. We used these spectra to
estimate the size of the Strontium Filament in \ion{Fe}{ii} and
[\ion{Sr}{ii}].

$^6$ Repeated spectra 4194-4593\AA; 4818-5104\AA\ to check variability
\end{table*}

Direct imagery of the \ion{Sr}~Filament is not possible through
the broad-bandpass filters available in the WFPC2 or the ACS
cameras. The relatively weak emission lines are overwhelmed by
dust-scattered starlight throughout the Homunculus. Other nebular
emission structures with different photo-excitations, different
spatial and velocity intervals are located in, or close to, the
line of sight towards the \ion{Sr}~Filament. With direct imagery
as in Figure 1
\citep{MDB98,SMG04}, we can trace the dusty
structures of the Homunculus by the scattered, red starlight.
Polarization measures using WFPC2 imagery \citep{SPC99,KNW02}
confirm the scattering properties of this light. The Little
Homunculus was detected by multiple emission lines of
\ion{Fe}{ii}, \ion{Cr}{ii}, etc in the near-ultraviolet, and may
contribute to the "purple haze", associated with the feature commonly 
called the 'Fan' within the Northwest lobe \citep{SMG04}.
Within the skirt, or the gas and dust structure located between
the two lobes, bright emission lines extend over significant
regions. Inspection of these emission lines indicate that some
emission extends, in velocity and space, around the
\ion{Sr}~Filament (Figure 2). The \ion{Sr}~Filament is best mapped
with high spatial and moderate spectral resolution as produced by
the STIS CCD moderate dispersion modes. Ground-based observations
are currently limited to half-arcsecond seeing with much scattered
light from the central star. The \ion{Sr}~Filament structure
then becomes confused with other nebular emission and stellar emission.

Spectra of the \ion{Sr}~Filament were recorded through HST visits
scheduled between February 1999 to December 2002 (2.8 years). This
extends over the mid-portion of $\eta$ Car's broad spectroscopic
maximum, which covers 5 years of the 5.52 year period. Highly
excited emission lines of [\ion{Ne}{iii}], [\ion{Ar}{iii}] and
\ion{He}{i}, as  monitored of the entire nebulosity by \citet{DSK98}, 
changed slowly over this interval in
nebular structure close to $\eta$ Car (but these lines are not 
detected in the \ion{Sr}~Filament). 

Given that these observations occur late in the broad spectroscopic maximum (when fluxes of 
all nebular lines appear to be relatively constant), we do not 
anticipate significant changes in the excitation of the Sr-Filament. The 
spectrum of the Weigelt Blobs, being at an order of magnitude closer to the 
Central Source, shows little change in the low-excitation emission lines of e.g.\ \ion{Fe}{ii}
 and \ion{Ni}{ii} across the entire 5.52-year cycle. Modeling by Verner et al (2002 and in 
preparation) demonstrates that the low-excitation emission lines are due largely 
to UV radiation longward of Lyman alpha. Most excitation of the Sr Filament 
appears to be due to mid-UV and near-UV, which changes little across the 
minimum. \citet{SMD00} investigated the photometric variability in the "purple haze" in 
WFPC2 pictures and found no evidence for variability that one might associate with the Sr filament,
even during the 5.5 year cycle and the brightening of the star.  

Where possible we repeated some spectral
overlap to check for variability in emission line fluxes. The
specific dates, HST programs, offsets from $\eta $ Car, position
angle, STIS aperture and spectral coverage are listed in Table 1.
Locations of the aperture positions are overlaid on an Advanced
Camera for Surveys (ACS) High Resolution Camera (HRC) ultraviolet
image of the Homunculus 
\citep{SMG04} in Figure 1.

The discovery spectrum was recorded on February 21, 1999 under
program 8036 \citep{ZGH01}. Under program 8327, deep spectra were
recorded on March 13, 2000 for two purposes: 1) we wanted deeper
exposures of the Homunculus to infer the spatial structure of the
lobes by the changes in velocity with position of locally-emitted
narrow nebular emission lines and the locally-absorbed narrow
absorption lines of \ion{Ca}{ii} \citep{DSG01}, and 2) we wanted
to extend the spectral coverage of the \ion{Sr}~Filament to other
spectral regions as we anticipated that emission lines from
additional elements in neutral and singly-ionized states would be
detected. As most emission lines in the Homunculus are marginally
resolved with the CCD G750M, G430M and G230MB gratings and the
$52\arcsec \times 0\farcs1$ aperture, a significant gain in
limiting flux was obtained by using the $ 52\arcsec \times 0\farcs
2$ aperture. For this visit with STIS, we  placed $\eta$ Car
behind the F2 fiducial ($0\farcs85$ wide), to prevent saturation
of the CCD by the very bright star. On March 21, 2000 (program
8483), the Homunculus was mapped with the STIS using the
$52\arcsec \times 0\farcs1$ aperture and the grating settings for
6480-7000\AA\ at $0\farcs 25$ spacing and for 4818-5100\AA\ at
$0\farcs 1$ spacing. The overall structure, with emphasis on the
internal ionized regions called the Little Homunculus, is discussed by
\citet{IGD03}. We used these same data to map the
\ion{Sr}~Filament. Deep spectra were recorded in April 2001
(program 8619) at wavelengths selected to obtain the fluxes of
\ion{Sr}{ii} emission lines near 4100\AA, to measure the width of
the \ion{Sr}~Filament, and to detect additional emission lines. On
November 27, 2001 (Program 8619), we recorded deep exposures from
3022-10135\AA\ and 2480-2633\AA\ as that was the region of the
spectrum where many emission lines were predicted, including the
\ion{Fe}{ii} lines at 2507\AA\ and 2509\AA. Based upon the
non-detection of the Lyman-alpha-pumped Fe II lines at 2507 and
2509\AA, we suspected that little or no hydrogen photo-ionizing
radiation was impinging upon the \ion{Sr}~Filament. We expected
that few emission lines would be detected below 2480\AA. On
December 16, 2002, we extended spectral coverage from 3052\AA\ to
1640\AA\ to gain full spectral coverage. No lines were detected
shortward of 2489.7\AA, which is an \ion{Fe}{i} line of multiplet uv9.
However, we caution the reader that the CCD sensitivity decreases
rapidly below 2500\AA. Deeper exposures would be possible with the
MAMA detectors, likely in the E230M mode.

\begin{figure*}
\begin{center}
\resizebox{\textwidth}{!}{\rotatebox{0}{\includegraphics{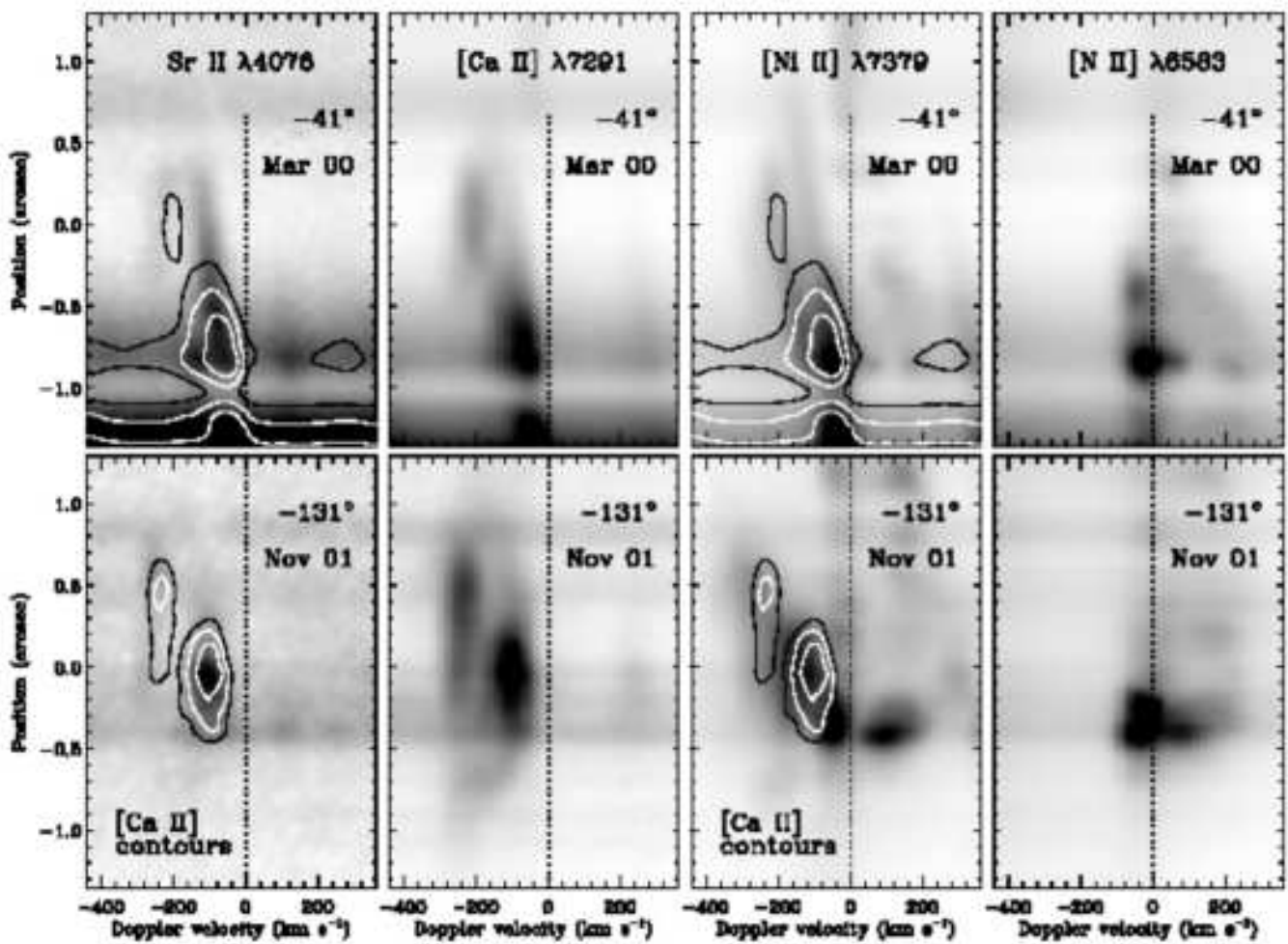}}}
\caption{Position-velocity diagrams at the location of the
strontium filament seen in several different emission lines, with
two different slit orientations.  The top row shows spectra taken
in March 2000 with the STIS slit at P.A.=-41$\degr$, and the bottom
row shows the same four emission lines obtained in November 2001,
with the slit at P.A.$=-131\degr$ (see Figure 1 and Table 1). The
four emission lines shown here are (left to right in both the top
and bottom panels): Sr~{\sc ii} $\lambda$4078, [Ca~{\sc ii}]
$\lambda$7293 (this is actually the average of [Ca~{\sc ii}]
$\lambda$7293 and [Ca~{\sc ii}] $\lambda$7325), [Ni~{\sc ii}]
$\lambda$7379, and [N~{\sc ii}] $\lambda$6585.  The contours
superposed on the Sr~{\sc ii} and [Ni~{\sc ii}] emission are
corresponding contours of the [Ca~{\sc ii}] emission at the same
dates and slit positions.  The horizontal axis shows heliocentric
Doppler velocity.  In the top and bottom panels, the position
marked as zero corresponds roughly to the place where the two
slits cross to within about 0$\farcs$1 (see Figure 1).\label{fig:sr2vel}}
\end{center}
\end{figure*}

\section{Kinematic structure of the strontium filament}
The structure of the strontium filament is complex, and varies
depending on the particular emission line observed.  Since several
distinct emission-line structures are seen projected along the
same line of sight, the morphology needs to be understood before
we can attempt to interpret the observed spectrum and assign line
identifications.  We know from previous study 
\citep{DSG01,S02,IGD03} that at the position of
the strontium filament there are at least three different
structures: the equatorial skirt, the receding northwest polar
lobe of the Homunculus (showing both intrinsic emission and
reflected light), and the Little Homunculus.  Fortunately, these
different features can be disentangled by considering their
Doppler shifts.

Figure 2 shows the kinematic structure at the position of the
strontium filament.  Two different position angles, perpendicular
to one another, are chosen to sample its full spatial extent (see
Figure 1).  The resonance line Sr~{\sc ii} $\lambda$4078 is the
brightest of the four lines from Sr$^+$ that have been detected in
our data \citep{ZGH01}. 
[Ca~{\sc ii}] is a much brighter
line that traces gas in a similar ionization state; Ca$^+$ ranges
from 6.1 to 11.9 eV, while Sr$^+$ has a range of 5.7 to 11.0 eV.
Indeed, contours of [Ca~{\sc ii}] superposed over Sr~{\sc ii}
emission show that both lines seem to trace the same
gas.\footnotemark\footnotetext{In regions exposed to radiation
above 11 to 12 eV, Sr and Ca will both be doubly ionized and
difficult to detect because of their atomic structure.} The
[Ca~{\sc ii}] emission shows that the \ion{Sr}~filament is more than
just a single thin ``filament'', and has a spatial extent of
$\ga$1$\farcs$5 along the polar axis of the Homunculus, and
$\ga$2$\arcsec$ in the direction perpendicular to the polar axis.
Thus, the \ion{Sr}~Filament occupies the same spatial extent as the
``purple haze'' seen in {\it HST} images of $\eta$ Car 
\citep{SMG04,MDB98}. The \ion{Sr}~Filament is constrained to
heliocentric velocities between $-$50 and $-$300 km s$^{-1}$, and
is probably in or near the equatorial plane.  It appears to have
two main velocity components; one at about $-$100 km s$^{-1}$,
with a velocity structure tilted so that emission becomes more
blueshifted with increasing separation from the star (in March
2000), and another feature at about $-$240 km s$^{-1}$ with the
opposite tilt.  The $-$100 km s$^{-1}$ component is brighter at
most positions (at least in Sr~{\sc ii}), and dominates the
emission spectrum listed in Table 2.  Interestingly, fainter
extended emission in [Ca~{\sc ii}] suggests that both these
velocity components may be part of a single closed structure,
forming a ring or loop in velocity space, especially in the
November 2001 spectra.

Emission from [Ni~{\sc ii}] $\lambda$7379, on the other hand,
shows subtle differences compared to both Sr~{\sc ii} and [Ca~{\sc
ii}].  It has two remarkably straight velocity components (top
panel in Figure 2), both tilted in the same sense, both with
blueshift increasing with separation from the star.  
\citet{DSG01} suggested that these two velocity components traced
gas in the equatorial plane with two different ages, originating
in the Great Eruption and the 1890 event \citep[see][]{HDS99}.
Some of the [Ni~{\sc ii}] emission coincides with Sr~{\sc
ii} and [Ca~{\sc ii}], but some does not.  In particular, the
diffuse [Ni~{\sc ii}] emission near position=0$\arcsec$ and $-$150
km s$^{-1}$ in the bottom panel of Figure 2 seems to fill in the
gap between the two velocity components of Sr~{\sc ii} and
[Ca~{\sc ii}].  Perhaps this makes sense, since Ni$^+$ occupies
ionization zones between 7.6 and 18.2 eV, only partly overlapping
with Sr$^+$ and Ca$^+$.  The velocity components of [Ni~{\sc ii}]
and [N~{\sc ii}] at $-$40 km s$^{-1}$ and redshifted velocities up
to $-$300 km s$^{-1}$ trace the northwest polar lobes of the
Homunculus and Little Homunculus, respectively 
\citep{IGD03,SMG04}. [N~{\sc ii}] emission is {\it only} seen in
these polar features, and is absent in the equatorial material,
while [Ni~{\sc ii}] is seen in both.  These polar structures must
be exposed to radiation above 12 eV, since Sr~{\sc ii} and
[Ca~{\sc ii}] are absent. This reinforces the idea that the
strontium filament is somehow shielded from radiation above 12 eV
(even though the ionization potential of N is 14.5 eV, it can be
ionized from the excited $^2$D state by photons at $\sim$12.1 eV).

As noted above, the blueshifted velocities imply that the
strontium filament may reside in or near the equatorial plane ---
this may be an important factor for understanding its unusual
excitation.  On the one hand, various clues suggest that $\eta$
Car has an asymmetric radiation field, with more UV radiation
escaping the stellar wind at low latitudes near the equator where
the wind is thinner \citep{SDG03,SMG04},
at least during its ``normal''
high-excitation state between spectroscopic events.  On the other
hand, there may be a considerable column of material between the
star and the strontium filament, including the Weigelt objects and
a larger-scale disk or torus, which may absorb all ionizing
photons along that path but apparently transmits photons below 12
eV.  In any case, both the strontium filament and the Weigelt
objects appear to occupy a special azimuthal direction relative to
$\eta$ Car.  In general, the subtle variations in emission
structure from one tracer to the next suggest that the strontium
filament is a low-ionization region with stratified ionization
zones.  This will be relevant in future efforts to model the
emission spectrum (Bautista et al., in preparation).

During the April 2001 observation, the slit was oriented in a way that a region NE of 
the filament (i.e.\ $\sim$2" north of the star) was observed. This spatial region close 
to the filament shows a weak scattered continuum. In this continuum can be 
seen absorption lines from the allowed \ion{Sr}{ii} lines $\lambda$4078,4216 and the 
\ion{Ca}{i} $\lambda$4227 line. For this slit orientation, there are only a few wavelength 
regions observed, but this absorption is not observed in any other lines. For 
all of the transitions showing this absorption, the lower level is the ground 
state, which might indicate that this is a low excitation region.
\begin{figure}
\resizebox{9cm}{!}{\includegraphics{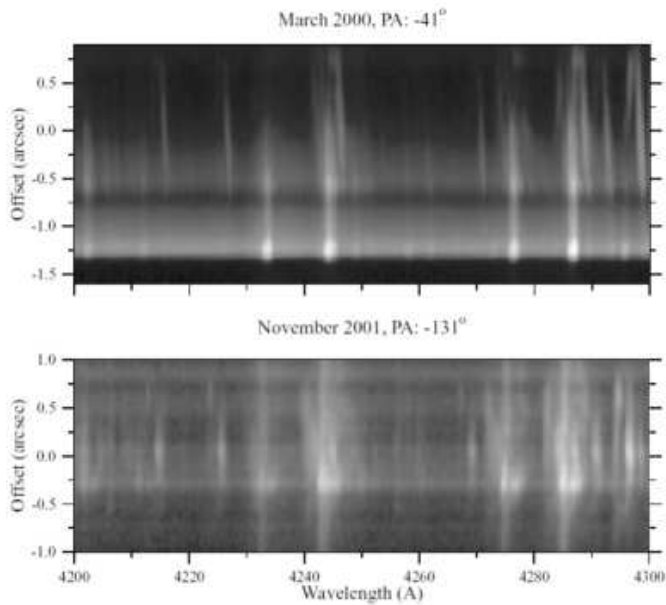}} 
\caption{2D spectrum of the filament at two different slit
orientations at nearly orthogonal angles, March 2000 and November
2001, respectively. 
The vertical stretch in the upper panel is 2$\farcs$5 and in the lower 2".
The horizontal scale covers the region
4200-4300\AA. The aperture width was 0."2 on the sky. 
The slit positions are shown projected on the
Homunculus in Fig. 1 with the plotted regions marked with bold 
lines. \label{diffslits}}
\end{figure}
\begin{figure}
\resizebox{9cm}{!}{\includegraphics{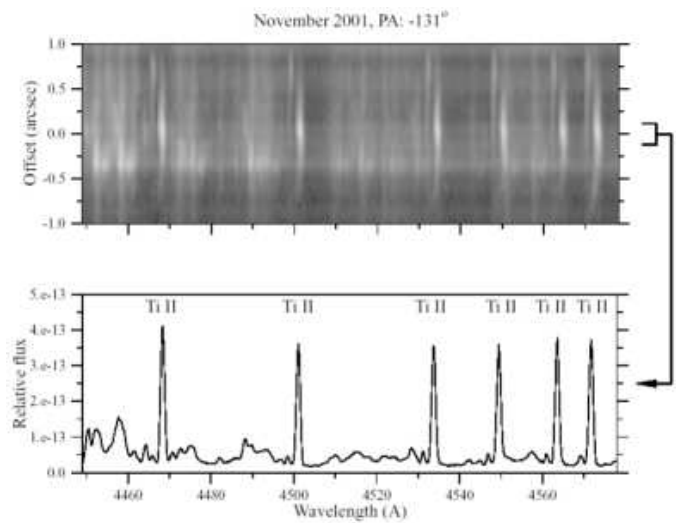}}
\caption{Upper panel: Subsection of STIS nebular spectrum centered
upon the \ion{Sr}~Filament . The STIS $52 \times 0\farcs2$
aperture was used for this spectrum. Lower panel: Extracted
spectrum of the $0\farcs2 \times 0\farcs2$ area, marked on the
right hand side of the nebular spectrum in the upper panel. This represents
the central position of the \ion{Sr}~Filament.\label{extrspec}}
\end{figure}
\section{Line Identification}
The spectra of the \ion{Sr}~Filament were observed at five
different dates with a long aperture at different position angles
as mentioned in Sec. 2 (See Figure 1 and Table 1). By inspection
of individual visits, we know that there are very significant
spatial variations of the nebular emission. Given the interval in
time between all observations, there is the possibility of
temporal variations. Other than in the region of close overlap,
there is also the strong possibility of spatial and/or temporal
variations complicating the application of line ratios, especially
if the relevant spectral lines were not measured in the same
visit. However, some spectral regions were observed on several
occasions. The different profiles for the same spectral line in
different visits can then be compared. We also note that the bulk
of the observations were taken thirteen months apart (in November
2001 and December 2002) at position angles within $16\degr$ of each
other. The star is in the broad, high-excitation phase;
temporal variability should be minimum across this time-span.
Spatial variations appear to be small in the central core of the
\ion{Sr}~Filament.

Some observed emission lines are likely the composite of emission
from a number of regions along line of sight. The imaging
properties of STIS along the aperture and the radial velocity
shifts in the spectral direction enable us to disentangle the
nebular emission for several kinematically different systems, 
as discussed in Sec. 3 above and seen in Fig. 2. Lines belonging 
to the filament show a Doppler shift of about -100~km/s at 
the center of the filament.

Examples of the spatial and velocity variations for two different
slit orientations are presented in Figure 3. The aperture position
angle was at $-41\degr$ (North through East convention) in March 2000
(upper plot) and is located in the general direction from $\eta$
Car along the polar axis of the Homunculus. In the 2$\farcs$5 slice (top, Figure 3),
the \ion{Sr}~Filament is centered at 0\arcsec. A velocity shift is
noticeable in the spectral dispersion to the blue at $0\farcs5$.
The November 2001 observations was with the aperture positioned at
$-131\degr$, or perpendicular to the Homunculus polar axis. The peak
of the \ion{Sr}~Filament emission is positioned at 0\arcsec. There
is little velocity shift along the slice of nebular in this
direction. In Figure 4, a second slice of spectrum is shown for
the November 2001 set of spectra (top). The extracted spectrum for
a $0\farcs2$-high extraction is plotted below. Note that all six
bright lines are \ion{Ti}{ii} emission lines of the -100km/s
\ion{Sr}~Filament. A much fainter second filament can be seen to
the blue in both the spectral image (top) and the extracted
spectrum(bottom). These two structures are plotted in velocity in
Figure 2.

\subsection{Linelist and intensities} \label{subsec:linelist}
All lines attributed to the \ion{Sr}~Filament are presented in
Table 2. The major selection criteria are the shape of the line
and its spatial location in the long-slit STIS spectrum (see Sec.
3). We do not observe any lines attributable to the
\ion{Sr}~Filament in the region below 2480~\AA. Little emission is
observed below 2000~\AA. Strong absorption along the line of sight
from singly-ionized iron-group elements removes much light at
these wavelengths, and the CCD sensitivity also declines in the
mid-UV region. The absence of lines could also be linked to the
possible presence of a lower wavelength limit for radiation
available for ionization and excitation. Some faint emission lines
are present but they may not be associable with the Sr Filament as emission 
lines are present from other regions along line of sight. 
The region 2500--3000~\AA\ is less affected by
absorption and shows emission lines. The region longward of
3000~\AA\ contains well-defined Sr Filament emission lines, although emission from
other spatial regions affects the spectrum.

The observed lines in Table 2 were measured in vacuo with
heliocentric velocity corrections (column 1) and, where
identified, include laboratory wavelengths (column 3). The
difference between the two is converted to a velocity (km/s,
column 2). The velocities are derived from the nov01 spectrum
except for a few cases, when the mar00 (2630-3025 \AA) spectra are
used. The spectra include emission lines from other nebular 
structures in line of sight, but only the lines associated with 
the strontium filament are
included in the linelist. Some lines have measured
velocities deviating from --100~km/s as they are either affected
by absorption or blended by other lines. Blended lines, or lines
with possible multiple identifications, are listed more than once
with the identified wavelength, but with alternate wavelengths
corresponding to the alternate line identifications. These lines
are associated with the strontium filament despite the deviant
Doppler velocity. The lines are identified by the species (e.g.
\ion{Fe}{i}), multiplet number and transition in columns 4-6. The
transition is represented by the lower and upper level, using the
$LS$ term notation in Moore's tables of multiplets and energy
levels. Thus, the term notation is preceded by a small letter
a,b,c,etc. for even parity configurations and z,y,x,etc. for
odd-parity configurations. If the multiplet is missing in Moore's
tables we have inserted an abbreviated configuration notation. For
full spectroscopic notations the reader is referred to the
original laboratory line lists or to detailed atomic databases
\footnote{\mbox{e.g.~http://physics.nist.gov/cgi-bin/AtData/main\_asd} 
\mbox{at NIST or the data by Kurucz at}\\
\mbox{http://cfa-www.harvard.edu/amdata/ampdata/kurucz23/sekur.html}}.
Intensities from the different STIS spectra are given in columns
7-10. Unidentified lines are marked with "unid" in the fourth
column.

In Table 3 we have sorted the identified lines from Table 2 according to
element, starting with the lightest element, carbon, and ending with the
heaviest, strontium. Within each species the lines have been grouped after the
excitation potential of the upper level, from which the line originates. The
velocity for each line has been included to facilitate the study of consistency
within each group of lines for a specific species.

\subsection{Observed lines}
The spectrum of the \ion{Sr}~Filament is dominated by permitted
and forbidden lines of the iron group elements, as can be seen in
Table 3. The spectrum is quite different from spectra of the
Weigelt blobs \citep{ZPHD01} as regards the line and intensity
distributions among the different elements. For example, lines of
\ion{Fe}{i}, \ion{V}{ii} and \ion{Ti}{ii} are particularly strong
in the \ion{Sr}~Filament, whereas only a few \ion{Fe}{ii} lines
are observed. Among the observed spectra are \ion{C}{i},
\ion{Mg}{i}, \ion{Al}{ii}, \ion{Ca}{i}, \ion{Ca}{ii},
\ion{Sc}{ii}, \ion{Cr}{ii}, \ion{Mn}{ii}, \ion{Fe}{ii},
\ion{Co}{ii}, \ion{Ni}{ii} and \ion{Sr}{ii}.
By contrast, lines of Fe II are dominant in the Weigelt blobs. \ion{Fe}{iii}
and \ion{Fe}{iv} lines are identified, but very few \ion{Fe}{i} lines \citep{ZPHD01}.
  
The spectral distribution of lines from a specific atom(ion) is determined by
the atomic structure and the value of the ionization potential. Hence, the
number of lines identified 
for different species reflects not only the abundance but also the 
complexity of the atomic structure.
Assuming an upper limit of about 8 eV of the photon energy
available for ionization and excitation (see Sec. 2) the number of
observable emission lines from some spectra will be very small. We
can divide the spectra of the observed species in three groups,
corresponding to their atomic structure and referring to the
periodic table:
\\ I) Group 1A and 2A (\ion{Na}{i}, \ion{Mg}{i}, \ion{Al}{ii}, \ion{Ca}{i},
\ion{Ca}{ii}, \ion{Sr}{ii})\\ II) Group 3A-7A (\ion{C}{i})\\
III) 3d-elements (\ion{Sc} - \ion{Ni})

Group I has quite simple spectra and the resonance lines appear in
the optical region for the alkali atoms. In alkali spectra there
are no forbidden spectral lines to observe, but the alkaline
earth-like ions \ion{Ca}{i} and \ion{Sr}{ii} show forbidden 3d-4s
(4d-5s) transitions due to the similar binding energy of d- and
s-electrons in transition group elements. In practice, these
elements could therefore be placed in group III, where the overlap
of 3d- and 4s configurations is the characteristic signature. In
general, the number of lines increases with the number of valence
electrons, i.e. the atomic number, for the transition elements,
but decreases towards the end of the period when the d-shell gets
closed. We see in Table 3 that there is a large number of lines
of \ion{Sc}{ii}, \ion{Ti}{ii} and \ion{V}{ii} compared to
\ion{Cr}{ii}, \ion{Mn}{ii} and \ion{Fe}{ii}. This excess of lines
for the first elements of the transition group could thus be an
abundance effect. However, we have to keep in mind that the
ionization energy, and therewith also the excitation energy for
the resonance lines increase with atomic number for the
iron-group. A limited photon energy for radiative excitation
favors the lightest iron-group elements, and the wavelengths of
the resonance lines is below 2480 \AA\ for the heavier ones.

\begin{figure}
\resizebox{9cm}{!}{\rotatebox{0}{\includegraphics{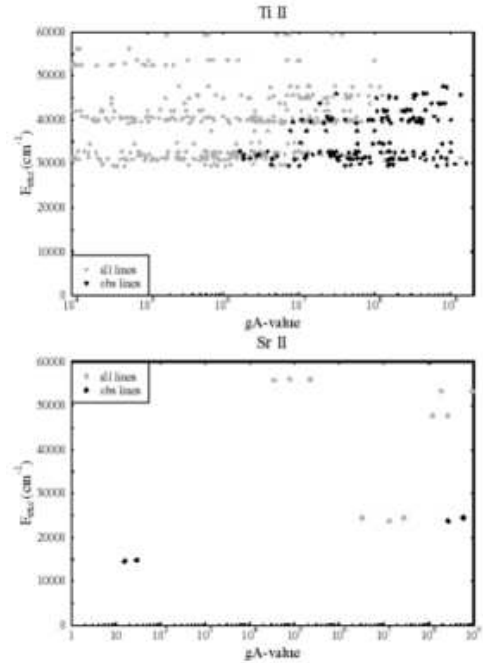}}}
\caption{Lines from \ion{Ti}{ii} (upper panel) and \ion{Sr}{ii} (lower panel) in
the region 3000-10000~Å for \ion{Ti}{ii} and 3000-11000 Å for \ion{Sr}{ii}. All
possible lines are shown in grey and the lines observed in the
spectrum are in black. The difference in atomic structure is
obvious. Note the different x-scales. The forbidden lines of
[\ion{Ti}{ii}] are not included.\label{TiIISrII}}
\end{figure}

Some single line identifications in Table 2 (and 3) are
questionable either because the associated velocity differs
remarkably from other lines or because the excitation energy is
much higher than for other lines of the same species. However, the
presence of the element is clear. As discussed above the presence
of elements having a simple atomic structure is difficult to
verify from the number of spectral lines. For example, aluminum is
detected only by the inter-combination line of \ion{Al}{ii} and
cannot be verified by other lines in the observed wavelength
region, assuming similar excitation energies as for other
elements.  For comparison, we show in Figure \ref{TiIISrII} the
situation for two spectra, \ion{Ti}{ii} and \ion{Sr}{ii},
where all allowed transitions within the observed wavelength range
are included. The Einstein A-value (times the statistical weight 
of the upper level) is given on the horizontal axis
as a measure of the line strength and the excitation energy of the
upper level of the transition on the vertical axis. Among all
possible transitions (marked with grey dots) the observed lines
are marked with black dots. An expected trend of decreasing level
population with increasing excitation energy can be seen. In
addition, we also clearly see the larger number of predicted lines
for \ion{Ti}{ii} due to a more complex atomic structure
(\ion{Ti}{ii} has three and \ion{Sr}{ii} one electron outside
closed shells).

No lines from hydrogen or helium that can be associated with
emission from the \ion{Sr}~Filament have been detected. Neither
have lines from nitrogen or oxygen. However, the spectrum includes
the dust-scattered hydrogen Balmer P-Cygni stellar emission and
could mask very weak nebular Balmer emission lines. The only
neutron-capture element identified is strontium, but two of the
unidentified lines, discussed in the next subsection, coincide in
wavelength with \ion{Y}{ii}. However, they cannot be confirmed by
other \ion{Y}{ii} lines having about the same probability to
occur. Two of the unidentified lines coincide with the two
strongest lines of the resonance multiplet of \ion{Zr}{ii},
a$^4$F-z$^4$G. Other than a coincidence with one other weak line, no
other lines from \ion{Zr}{ii} are observed.

The line intensities from individual observations are included in
Table 2. They are represented by the integrated flux in the
emission feature, where contributions from obvious blending
components have been subtracted. The tabulated flux is measured
from intensity-calibrated spectra, which have not been corrected
for interstellar reddening. Since the different observations do
not cover the same spatial region the intensity ratios for
different line pairs may not be the same in the different
observations. In the wavelength regions affected by foreground
absorption the intensity values are less reliable. Generally, the
accuracy of the intensities is also affected by blending from
other lines as well as from emission from the same line formed in
other spatial regions along the line of sight. Some lines of e.g.\
[\ion{Fe}{ii}] and [\ion{Ni}{ii}] show complex profiles. The
intensity contribution from the \ion{Sr}~Filament is difficult to
determine. In such cases we  used  multi-Gaussian fits to extract
the fluxes.

\subsection{Unidentified lines}
About 40 of the 600 observed lines in the spectra of the
\ion{Sr}~Filament remain unidentified. These are marked with
$unid$ in Table 2 and listed in Table 4. A dozen unidentified
lines have  substantial strengths, as indicated in the intensity
column of Table 4. For some of the lines we include possible
identifications, but they should only be regarded as wavelength
coincidences between observed lines and predicted transitions. The
reasons for not including them among the identified lines could be
a large Doppler velocity, an anomalous excitation energy or a
general inconsistency with observed lines from the same ion.

Perhaps the most striking features of Table 4 are the wavelength
coincidences of two \ion{Y}{ii} lines and of two \ion{Zr}{ii} lines
and also the absence of candidates for identification of three of the strongest lines.
The latter three lines appear below 2600 \AA, which means that the
corresponding photon energy is about 4.8 eV. This is remarkable,
considering that 4.8 eV is not far from the highest excitation
energy observed in the total spectrum and that all transitions
between levels below 7-8 eV are known in ionized iron-group
elements. The most probable explanation for these lines is that
they originate from a neutral atom. However, as mentioned in
section \ref{subsec:linelist}, this region suffers from absorption
which can cause a shift in the observed wavelength and a
significantly decreased measured line intensity. Emission from
other spatial regions can also affect the lines.

\section{Discussion}
The \ion{Sr}~Filament proves to be a very unusual emission
nebulosity. Over 600 emission lines, mostly from neutral and
singly-ionized iron-peak elements, have been identified. Yet no
hydrogen, helium, nitrogen or oxygen emission lines,
which characterize normal emission regions, have been detected.
Several factors contribute to this unlikely emission-nebula
spectrum: 1) the very massive star  system, while producing
$5\times10^6 L_{\odot}$ with a characteristic temperature of
$25,000 \degr K$ \citep{HDI01}, has a highly clumpy, extended
but cool atmosphere \citep{VBV03}; 2)
much ionized gas shields this \ion{Sr}~Filament from hard UV
radiation capable of ionizing hydrogen, and other elements
requiring photo-ionizing energies exceeding 13.6eV; 3) apparently
the shielding may be sufficient to protect the \ion{Sr}~Filament
from radiation down to 6 or 7 eV as we see no direct evidence of
line emission requiring photo-excitation with that energy; 4) an
intense radiation field with energies less then 6 or 7 eV does
bathe the Sr Filament, leading to a partially-ionized region; 5)
the \ion{Sr}{ii} modelling \citep{BGI02} indicates a
very high-density region as the electron density must be in the
range of $10^7$ cm$^{-3}$.

We note that the ejecta surrounding $\eta$ Car have a very
non-uniform structure. While the overall Homunculus is a thin,
hollow shell about ten percent thickness compared to the distance
from the Central Source \citep{S02,SGH03} the interior is likely a 
hot, low-density stellar wind. The thin surface interior to the
shell is detected in [\ion{Fe}{ii}], [\ion{Ni}{ii}] emissions and \ion{Ca}{ii}
absorption \citep{DSG01,S02}. In line of sight, ejecta absorptions
of the iron-peak elements demonstrate a range of temperature and
electron density that correlate with velocity (Gull, et al, \apj\
submitted). Interior to the Homunculus is the Little Homunculus, a
miniature bilobed structure \citep{IGD03}, which is seen in
multiple emission lines of iron-peak elements and in the hydrogen
Balmer lines. Between the bi-lobes of the Homunculus and the
Little Homunculus is the skirt region, partially seen in emission
lines, and in absorption lines.
Close to $\eta$ Car are several very intense emission knots, the
Weigelt Blobs, seen strongly in many \ion{Fe}{ii}, \ion{Ni}{ii}, and \ion{Cr}{ii} lines
\citep{ZPHD01}. Highly-excited emission lines seen in the Weigelt
Blobs and in the Little Homunculus disappeared during the 1998.0
and 2003.5 minima, but then returned. 
Low-excitation lines maintain constancy in flux throughout the 5.52-year period. 
As the Sr Filament  emission lines are low-excitation only, this reinforces the 
concept that the \ion{Sr}~Filament receives
radiation with the harder photons filtered out. Based upon the
spatial distribution of the blue-shifted velocities, the
\ion{Sr}~Filament is probably located in this equatorial skirt
region. At a projected distance of the order of ten light-days,
the \ion{Sr}~Filament receives intense mid-UV, and longer
wavelength, radiation from the Central Source. Likely it is the
strong absorption by iron and other iron-peak elements in the
ionized regions and just beyond the ionized regions that shield
the \ion{Sr}~Filament. We note that the ionization potential of Fe
is 7.9 eV and that of Sr is 5.7 eV. As there is abundant
\ion{Fe}{i}  and little \ion{Fe}{ii} in the \ion{Sr}~Filament,
this indicates that the strontium is singly-ionized, but protected
from becoming doubly-ionized by an iron-shield. Moreover, many
\ion{Fe}{ii} absorptions are in the spectrum indicating the
abundance of singly-ionized iron in the vicinity of the
\ion{Sr}~Filament. Shortward of 2500\AA, much of the ultraviolet
spectrum is chopped up, further protecting the neutral and
singly-ionized species with ionization potentials above 4 or 5
electron volts. Indeed the question arises as to whether molecular
species might reside in this region. Ground-based, near-IR
observations of the Homunculus \citep{S02} do not
indicate molecular hydrogen at these velocities or spatial
position, but the ground-based observations were accomplished with
lower spatial resolution.

We have systematically obtained spectra of the \ion{Sr}~Filament
to characterize the spatial extent and the level of excitation
through the identification of over 600 emission lines. We have
also measured the fluxes of these lines in preparation for
obtaining physical information of this neutral emission region.
The first paper, based upon the \ion{Sr}{ii} emission line ratios,
has been published, characterizing the temperature and density of
the \ion{Sr}~Filament \citep{BGI02}, other papers will follow
discussing models of other iron-peak neutral and singly-ionized
species. We hope to provide information on relative abundances of
various ionic species, possibly elemental abundances, but
modelling and possibly some laboratory work will first be
necessary.

\begin{acknowledgements}
We are grateful to Kazunori Ishibashi for providing calibrated spectra and
giving inputs in the initial analysis. Other members of the HST 
Eta~Carinae Treasury Project Team are 
Manuel Bautista, Michael Corcoran, Augusto Damineli, Kris Davidson (P.I.), 
Fred Hamann, John Hillier, Roberta Humphreys, Jon Morse, Otmar Stahl, 
Nolan Walborn and Kerstin Weis.
This study is part of a project funded through a contract (S.J.)
with the Swedish National Space Board. The data were obtained
through the following HST observational programs: 8036, 8327,
8483, 8619 and 9420. Funding was provided under the STIS GTO
program and HST GO programs. We acknowledge the assistance on the
analysis by the STIS Instrument Definition Team (IDT), especially
Don Lindler, Terry Beck and Keith Feggans. H.H.\ is grateful
for travel support from the STIS IDT for a visit to Goddard Space
Flight Center.  N.S.\ was supported by NASA through grant
HF-01166.01A from the Space Telescope Science Institute, which is
operated by the Association of Universities for Research in
Astronomy, Inc., under NASA contract NAS 5-26555. This research
has made use of NASA's Astrophysics Data System Bibliographic
Services.
\end{acknowledgements}

\bibliographystyle{aa}
\bibliography{../bibtex/hartman}
\renewcommand{\baselinestretch}{1}
\onecolumn
\begin{footnotesize}
\begin{table}
\caption{Spectral lines observed in the Sr-filament of $\eta$~Car in the wavelength region 2480-10140~\AA. The lines are sorted by wavelength.}
\begin{tabular}{ccccccccccc} \hline \hline
$\lambda_{obs}$	&	Velocity	&	$\lambda_{lab}$	&	Ion &	Mult.	&	Transition$^a$ &	I(mar00)	&	I(apr01)&	I(nov01)	&	I(dec02)		&	Comment$^b$	\\
(\AA)	&	(km/s)	&	(\AA)	&		&		&		&		\multicolumn{4}{c}{(10$^{-15}$ erg cm$^{-2}$ s$^{-1}$ arcsec$^{-2}$)}						&		\\ \hline
2489.71	&	-95	&	2490.501	&	FeI	&	(uv9)	&	a$^5$D$_0$-x$^5$F$_1$        &		&		&		&		&		\\        
2490.61	&	-95	&	2491.395	&	FeI	&	(uv9)	&	a$^5$D$_2$-x$^5$F$_3$        &		&		&	20	&		&		\\        
2491.05	&	-103	&	2491.907	&	FeI	&	(uv9)	&	a$^5$D$_1$-x$^5$F$_2$        &		&		&	4	&		&		\\        
2492.53	&	-103	&	2493.383	&	FeI	&	(uv63)	&	a$^5$F$_3$-x$^3$G$_3$        &		&		&		&		&		\\        
2492.82	&	-91	&	2493.573	&	FeI	&	(uv59)	&	a$^5$F$_3$-w$^5$G$_2$        &		&		&	22	&		&		\\        
2510.71	&	-105	&	2511.591	&	FeI	&	(uv7)	&	a$^5$D$_3$-x$^5$D$_2$        &		&		&	45	&		&		\\
\hline     
\end{tabular}
\end{table}
\addtocounter{table}{-1}
\begin{table}
\caption{Continued.}
\begin{tabular}{ccccccccccc} \hline \hline
$\lambda_{obs}$	&	Velocity	&	$\lambda_{lab}$	&	Ion &	Mult.	&	Transition$^a$ &	I(mar00)	&	I(apr01)&	I(nov01)	&	I(dec02)		&	Comment$^b$	\\
(\AA)	&	(km/s)	&	(\AA)	&		&		&		&		\multicolumn{4}{c}{(10$^{-15}$ erg cm$^{-2}$ s$^{-1}$ arcsec$^{-2}$)}						&		\\ \hline
2739.34	&	-112	&	2740.359	&	FeII	&	(uv63)	&	a$^4$D$_{7/2}$-z$^4$D$_{7/2}$    &	42	&		&		&		&		\\        
2743.12	&	-97	&	2744.008	&	FeII	&	(uv62)	&	a$^4$D$_{1/2}$-z$^4$F$_{3/2}$    &	35	&		&		&		&		\\        
2746.19	&	-121	&	2747.296	&	FeII	&	(uv62)	&	a$^4$D$_{1/2}$-z$^4$F$_{5/2}$    &		&		&		&		&		\\        
2749.03	&	-120	&	2750.134	&	FeII	&	(uv62)	&	a$^4$D$_{5/2}$-z$^4$F$_{7/2}$    &	67	&		&		&		&		\\        
2749.03	&	-138	&	2750.299	&	FeII	&	(uv63)	&	a$^4$D$_{1/2}$-z$^4$D$_{1/2}$    &	67	&		&		&		&		\\        
2755.54	&	-110	&	2756.552	&	FeII	&	(uv62)	&	a$^4$D$_{7/2}$-z$^4$F$_{9/2}$    &	128	&		&		&		&		\\
\hline      
\end{tabular}
\end{table}
\addtocounter{table}{-1}
\begin{table}
\caption{Continued.}
\begin{tabular}{ccccccccccc} \hline \hline
$\lambda_{obs}$	&	Velocity	&	$\lambda_{lab}$	&	Ion &	Mult.	&	Transition$^a$ &	I(mar00)	&	I(apr01)&	I(nov01)	&	I(dec02)		&	Comment$^b$	\\
(\AA)	&	(km/s)	&	(\AA)	&		&		&		&		\multicolumn{4}{c}{(10$^{-15}$ erg cm$^{-2}$ s$^{-1}$ arcsec$^{-2}$)}						&		\\ \hline
2968.46	&	-79	&	2969.241	&	VII	&	(28)	&	a$^5$P$_3$-y$^5$D$_4$        &		&		&		&	18	&		\\        
2970.42	&	-97	&	2971.382	&	FeII	&	(2)	&	a$^4$D$_{3/2}$-z$^6$F$_{5/2}$    &		&		&		&	35	&		\\        
2979.54	&	-69	&	2980.224	&	FeII	&	(2)	&	a$^4$D$_{1/2}$-z$^6$F$_{3/2}$    &		&		&		&	28	&	bl.	\\        
2994.41	&	-89	&	2995.300	&	FeI	&	(9)	&	a$^5$D$_3$-y$^5$D$_2$        &		&		&		&	8	&		\\        
2995.96	&	-91	&	2996.863	&	VII	&	(27)	&	a$^5$P$_1$-z$^5$P$_2$        &		&		&		&	7	&		\\        
2999.49	&	-90	&	3000.387	&	FeI	&	(30)	&	a$^5$F$_5$-x$^5$F$_5$        &		&		&		&	7	&		\\
\hline
\end{tabular}
\end{table}
\addtocounter{table}{-1}
\begin{table}
\caption{Continued.}
\begin{tabular}{ccccccccccc} \hline \hline
$\lambda_{obs}$	&	Velocity	&	$\lambda_{lab}$	&	Ion &	Mult.	&	Transition$^a$ &	I(mar00)	&	I(apr01)&	I(nov01)	&	I(dec02)		&	Comment$^b$	\\
(\AA)	&	(km/s)	&	(\AA)	&		&		&		&		\multicolumn{4}{c}{(10$^{-15}$ erg cm$^{-2}$ s$^{-1}$ arcsec$^{-2}$)}						&		\\ \hline
3120.09	&	-113	&	3121.264	&	CrII	&	(5)	&	a$^4$D$_{3/2}$-z$^4$F$_{5/2}$    &		&		&	34	&		&		\\        
3120.09	&	-62	&	3120.730	&	TiII	&	(67)	&	b$^4$P$_{5/2}$-z$^4$P$_{3/2}$    &		&		&	34	&		&		\\        
3120.09	&	-187	&	3122.041	&	VII	&	(1)	&	a$^5$F$_5$-z$^5$G$_5$        &		&		&	34	&		&		\\        
3124.94	&	-90	&	3125.879	&	CrII	&	(5)	&	a$^4$D$_{5/2}$-z$^4$F$_{7/2}$    &		&		&	58	&		&		\\        
3124.94	&	-119	&	3126.182	&	VII	&	(1)	&	a$^5$F$_1$-z$^5$G$_2$        &		&		&	58	&		&		\\        
3124.94	&	-209	&	3127.117	&	VII	&	(1)	&	a$^5$F$_4$-z$^5$G$_4$        &		&		&	58	&		&		\\
\hline      
\end{tabular}
\end{table}
\addtocounter{table}{-1}
\begin{table}
\caption{Continued.}
\begin{tabular}{ccccccccccc} \hline \hline
$\lambda_{obs}$	&	Velocity	&	$\lambda_{lab}$	&	Ion &	Mult.	&	Transition$^a$ &	I(mar00)	&	I(apr01)&	I(nov01)	&	I(dec02)		&	Comment$^b$	\\
(\AA)	&	(km/s)	&	(\AA)	&		&		&		&		\multicolumn{4}{c}{(10$^{-15}$ erg cm$^{-2}$ s$^{-1}$ arcsec$^{-2}$)}						&		\\ \hline
3278.15	&	-157	&	3279.868	&	TiII	&	(24)	&	a$^2$D$_{5/2}$-z$^2$P$_{3/2}$    &		&		&	60	&		&		\\        
3282.06	&	-111	&	3283.271	&	TiII	&	(66)	&	b$^4$P$_{1/2}$-y$^4$D$_{1/2}$    &		&		&	55	&		&		\\        
3287.72	&	-81	&	3288.608	&	TiII	&	(89)	&	b$^2$G$_{7/2}$-z$^2$H$_{9/2}$    &		&		&	47	&		&		\\        
3287.72	&	-165	&	3289.531	&	TiII	&	(66)	&	b$^4$P$_{3/2}$-y$^4$D$_{1/2}$    &		&		&	47	&		&		\\        
3295.98	&	-72	&	3296.766	&	FeII	&	(1)	&	a$^4$D$_{3/2}$-z$^6$D$_{3/2}$    &		&		&	14	&		&		\\        
3298.60	&	-99	&	3299.684	&	VII	&	(7)	&	a$^3$F$_4$-z$^3$G$_4$        &		&		&	5	&		&		\\
\hline
\end{tabular}
\end{table}
\addtocounter{table}{-1}
\begin{table}
\caption{Continued.}
\begin{tabular}{ccccccccccc} \hline \hline
$\lambda_{obs}$	&	Velocity	&	$\lambda_{lab}$	&	Ion &	Mult.	&	Transition$^a$ &	I(mar00)	&	I(apr01)&	I(nov01)	&	I(dec02)		&	Comment$^b$	\\
(\AA)	&	(km/s)	&	(\AA)	&		&		&		&		\multicolumn{4}{c}{(10$^{-15}$ erg cm$^{-2}$ s$^{-1}$ arcsec$^{-2}$)}						&		\\ \hline
3461.30	&	-103	&	3462.486	&	TiII	&	(6)	&	b$^4$F$_{7/2}$-z$^4$G$_{9/2}$    &		&		&	8	&		&		\\        
3465.65	&	-105	&	3466.853	&	FeI	&	(6)	&	a$^5$D$_1$-z$^5$P$_1$        &		&		&	6	&		&		\\        
3473.77	&	-110	&	3475.035	&	MnII	&	(3)	&	a$^5$D$_3$-z$^5$P$_3$        &		&		&	71	&		&		\\        
3473.77	&	-117	&	3475.124	&	MnII	&	(3)	&	a$^5$D$_2$-z$^5$P$_1$        &		&		&	71	&		&		\\        
3477.04	&	-99	&	3478.183	&	TiII	&	(6)	&	b$^4$F$_{5/2}$-z$^4$G$_{7/2}$    &		&		&	44	&		&		\\        
3482.53	&	-118	&	3483.902	&	MnII	&	(3)	&	a$^5$D$_2$-z$^5$P$_2$        &		&		&	64	&		&		\\
\hline
\end{tabular}
\end{table}
\addtocounter{table}{-1}
\begin{table}
\caption{Continued.}
\begin{tabular}{ccccccccccc} \hline \hline
$\lambda_{obs}$	&	Velocity	&	$\lambda_{lab}$	&	Ion &	Mult.	&	Transition$^a$ &	I(mar00)	&	I(apr01)&	I(nov01)	&	I(dec02)		&	Comment$^b$	\\
(\AA)	&	(km/s)	&	(\AA)	&		&		&		&		\multicolumn{4}{c}{(10$^{-15}$ erg cm$^{-2}$ s$^{-1}$ arcsec$^{-2}$)}						&		\\ \hline
3684.99	&	-102	&	3686.238	&	TiII	&	(14)	&	a$^2$F$_{5/2}$-z$^2$D$_{3/2}$    &		&		&	33	&		&		\\        
3684.99	&	-103	&	3686.253	&	TiII	&	(14)	&	a$^2$F$_{7/2}$-z$^2$D$_{5/2}$    &		&		&	33	&		&		\\        
3687.65	&		&		&	unid	&		&	                         &		&		&	11	&		&		\\        
3697.79	&		&		&	unid	&		&	                         &		&		&	4	&		&		\\        
3706.07	&	-98	&	3707.271	&	TiII	&	(72)	&	b$^2$D$_{3/2}$-y$^2$D$_{3/2}$    &		&		&	18	&		&		\\        
3709.69	&	-50	&	3710.301	&	FeI	&	(21)	&	a$^5$F$_4$-y$^5$F$_3$        &		&		&	9	&		&	id?	\\
\hline
\end{tabular}
\end{table}
\addtocounter{table}{-1}
\begin{table}
\caption{Continued.}
\begin{tabular}{ccccccccccc} \hline \hline
$\lambda_{obs}$	&	Velocity	&	$\lambda_{lab}$	&	Ion &	Mult.	&	Transition$^a$ &	I(mar00)	&	I(apr01)&	I(nov01)	&	I(dec02)		&	Comment$^b$	\\
(\AA)	&	(km/s)	&	(\AA)	&		&		&		&		\multicolumn{4}{c}{(10$^{-15}$ erg cm$^{-2}$ s$^{-1}$ arcsec$^{-2}$)}						&		\\ \hline
3903.05	&	-102	&	3904.374	&	VII	&	(11)	&	a$^3$P$_2$-z$^5$D$_3$        &	7	&		&	9	&		&		\\        
3913.27	&	-100	&	3914.576	&	TiII	&	(34)	&	a$^2$G$_{7/2}$-z$^2$G$_{7/2}$    &	23	&		&	31	&		&		\\        
3916.19	&	-102	&	3917.519	&	VII	&	(10)	&	a$^3$P$_1$-z$^3$D$_2$        &	5	&		&	7	&		&		\\        
3932.39	&	-182	&	3934.777	&	CaII	&	(1)	&	4s~$^2$S$_{1/2}$-4p~$^2$P$_{3/2}^o$    &	10	&		&	19	&		&		\\        
3951.84	&	-94	&	3953.083	&	VII	&	(10)	&	a$^3$P$_2$-z$^3$D$_3$        &	5	&		&	8	&		&		\\        
3958.07	&		&		&	unid	&		&	                         &	1	&		&	1	&		&		\\
\hline
\end{tabular}
\end{table}
\addtocounter{table}{-1}
\begin{table}
\caption{Continued.}
\begin{tabular}{ccccccccccc} \hline \hline
$\lambda_{obs}$	&	Velocity	&	$\lambda_{lab}$	&	Ion &	Mult.	&	Transition$^a$ &	I(mar00)	&	I(apr01)&	I(nov01)	&	I(dec02)		&	Comment$^b$	\\
(\AA)	&	(km/s)	&	(\AA)	&		&		&		&		\multicolumn{4}{c}{(10$^{-15}$ erg cm$^{-2}$ s$^{-1}$ arcsec$^{-2}$)}						&		\\ \hline
4383.33	&	-99	&	4384.776	&	FeI	&	(41)	&	a$^3$F$_4$-z$^5$G$_5$        &		&	17	&	14	&		&		\\        
4394.80	&	-100	&	4396.268	&	TiII	&	(19)	&	a$^2$D$_{5/2}$-z$^2$F$_{7/2}$    &		&	46	&	43	&		&		\\        
4399.80	&	-83	&	4401.008	&	TiII	&	(51)	&	a$^2$P$_{3/2}$-z$^4$D$_{5/2}$    &		&	29	&	27	&		&		\\        
4399.80	&	-125	&	4401.625	&	ScII	&	(14)	&	a$^3$F$_3$-z$^3$F$_3$        &		&		&	27	&		&		\\        
4404.65	&	-91	&	4405.987	&	FeI	&	(41)	&	a$^3$F$_3$-z$^5$G$_4$        &		&	8	&	6	&		&		\\        
4415.16	&	-82	&	4416.362	&	FeI	&	(41)	&	a$^3$F$_2$-z$^5$G$_3$        &		&	50	&	44	&		&		\\
\hline
\end{tabular}
\end{table}
\addtocounter{table}{-1}
\begin{table}
\caption{Continued.}
\begin{tabular}{ccccccccccc} \hline \hline
$\lambda_{obs}$	&	Velocity	&	$\lambda_{lab}$	&	Ion &	Mult.	&	Transition$^a$ &	I(mar00)	&	I(apr01)&	I(nov01)	&	I(dec02)		&	Comment$^b$	\\
(\AA)	&	(km/s)	&	(\AA)	&		&		&		&		\multicolumn{4}{c}{(10$^{-15}$ erg cm$^{-2}$ s$^{-1}$ arcsec$^{-2}$)}						&		\\ \hline
5494.38	&	-105	&	5496.302	&	[MnII]	&	(9F)	&	a$^5$D$_3$-b$^5$D$_4$        &		&		&	7	&		&		\\        
5526.53	&	-97	&	5528.325	&	ScII	&	(31)	&	a$^1$G$_4$-z$^1$F$_3$        &		&		&	14	&		&		\\        
5559.85	&		&		&	unid	&		&	                         &		&		&	4	&		&		\\        
5599.35	&		&		&	unid	&		&	                         &		&		&	1	&		&		\\        
5639.42	&	-91	&	5641.119	&	[FeI]	&	(2F)	&	a$^5$D$_4$-a$^5$P$_2$        &		&		&	12	&		&		\\        
5639.42	&	-168	&	5642.567	&	ScII	&	(29)	&	a$^3$P$_1$-z$^3$P$_2$        &		&		&	12	&		&	bl.	\\
\hline    
\end{tabular}
\end{table}
\addtocounter{table}{-1}
\begin{table}
\caption{Continued.}
\begin{tabular}{ccccccccccc} \hline \hline
$\lambda_{obs}$	&	Velocity	&	$\lambda_{lab}$	&	Ion &	Mult.	&	Transition$^a$ &	I(mar00)	&	I(apr01)&	I(nov01)	&	I(dec02)		&	Comment$^b$	\\
(\AA)	&	(km/s)	&	(\AA)	&		&		&		&		\multicolumn{4}{c}{(10$^{-15}$ erg cm$^{-2}$ s$^{-1}$ arcsec$^{-2}$)}						&		\\ \hline
7331.89	&	-89	&	7334.077	&	[VII]	&	(4F)	&	a$^5$D$_2$-a$^5$P$_3$        &	7	&	7	&	8	&		&		\\        
7343.90	&	-87	&	7346.030	&	[VII]	&	(12F)	&	a$^5$F$_3$-b$^3$G$_4$        &		&	2	&	2	&		&		\\        
7353.97	&	-74	&	7355.792	&	[VII]	&	(4F)	&	a$^5$D$_0$-a$^5$P$_2$        &	3	&	4	&	5	&		&		\\        
7377.22	&	-112	&	7379.970	&	[NiII]	&	(2F)	&	a$^2$D$_{5/2}$-a$^2$F$_{7/2}$    &	188	&	170	&	213	&		&		\\        
7411.22	&	-100	&	7413.690	&	[NiII]	&	(2F)	&	a$^2$D$_{3/2}$-a$^2$F$_{5/2}$    &	109	&	83	&	98	&		&		\\        
7426.27	&		&		&	unid	&		&	                         &	4	&	5	&	7	&		&		\\
\hline
\end{tabular}
\end{table}
\addtocounter{table}{-1}
\begin{table}
\caption{Continued.}
\begin{tabular}{ccccccccccc} \hline \hline
$\lambda_{obs}$	&	Velocity	&	$\lambda_{lab}$	&	Ion &	Mult.	&	Transition$^a$ &	I(mar00)	&	I(apr01)&	I(nov01)	&	I(dec02)		&	Comment$^b$	\\
(\AA)	&	(km/s)	&	(\AA)	&		&		&		&		\multicolumn{4}{c}{(10$^{-15}$ erg cm$^{-2}$ s$^{-1}$ arcsec$^{-2}$)}						&		\\ \hline
8726.82	&	-93	&	8729.523	&	[CI]	&	(3F)	&	2p$^2$~$^1$D$_2$-2p$^2$~$^1$S$_0$        &		&		&	15	&		&		\\        
8762.84	&	-97	&	8765.679	&	[VII]	&	(2F)	&	a$^5$D$_2$-a$^3$P$_1$        &		&		&	7	&		&		\\        
8806.31	&		&		&	unid	&		&	                         &		&		&	4	&		&		\\        
8878.60	&	-94	&	8881.373	&	[VII]	&	(2F)	&	a$^5$D$_1$-a$^3$P$_0$        &		&		&	5	&		&		\\        
9334.98	&	-110	&	9338.400	&	[CoII]	&	(1F)	&	a$^3$F$_4$-b$^3$F$_3$        &	2	&		&	6	&		&		\\        
9341.82	&		&		&	unid	&		&	                         &	1	&		&	4	&		&		\\        
9405.00	&	-107	&	9408.356	&	[TiII]	&	(21F)	&	a$^2$F$_{5/2}$-b$^2$G$_{9/2}$    &	6	&		&	3	&		&		\\        
9638.81	&	-95	&	9641.850	&	[CoII]	&	(1F)	&	a$^3$F$_3$-b$^3$F$_2$        &	4	&		&	4	&		&		\\        
9642.18	&	-90	&	9645.089	&	[TiII]	&	(21F)	&	a$^2$F$_{7/2}$-b$^2$G$_{7/2}$    &	1	&		&	0.4	&		&		\\        
9649.74	&	-92	&	9652.716	&	[TiII]	&	(21F)	&	a$^2$F$_{7/2}$-b$^2$G$_{9/2}$    &	1	&		&	2	&		&		\\        
9824.56	&	-69	&	9826.822	&	[CI]	&	(1F)	&	2p$^2$~$^3$P$_1$-2p$^2$~$^1$D$_2$        &	0.3	&		&	1	&		&		\\        
9850.24	&	-83	&	9852.965	&	[CI]	&	(1F)	&	2p$^2$~$^3$P$_2$-2p$^2$~$^1$D$_2$        &	2	&		&	2	&		&		\\        
10066.4	&	-100	&	10069.740	&	[TiII]	&	(5F)	&	a$^4$F$_{5/2}$-b$^4$P$_{5/2}$    &	5	&		&	3	&		&		\\        
10066.4	&	-103	&	10069.850	&	[TiII]	&	(5F)	&	a$^4$F$_{3/2}$-b$^4$P$_{3/2}$    &	5	&		&	3	&		&		\\        
10125.73	&	-90	&	10128.770	&	[TiII] & (5F)	&	a$^4$F$_{3/2}$-b$^4$P$_{1/2}$    &	5	&		&	7	&		&		\\         \hline
\end{tabular}\\
\hspace{-\parindent}$^a$ The configurations in complex spectra are replaced by symbols using the system introduced by \citet{M45}. For each ion the lowest even parity $LS$-term of each type is assigned the prefix a, the next b, and so on. Similarily, the odd terms are assigned the prefixes z,y,x etc. For example, the lowest even $^4$D in \ion{Fe}{ii} is written as a$^4$D, the second b$^4$D and the lowest odd $^4$D is marked z$^4$D. This system is not applied to the light elements, where the odd terms are marked with the superscript $^o$, for example $^3$P$_1^o$.\\ 
$^b$ Abbreviations used in the comment column:\\ 
bl.=lines blended by features not included in the list.\\
bl.abs=lines being partially blended by an absorption feature.\\
bl.HP=lines affected by hot pixels.\\
id?=identification uncertain.\\
Filament lines with more than one plausible identification are listed with the same measured wavelength, $\lambda_{obs}$, in column 1 but are not explicitly marked as blends in the Comment column.\\
\end{table}
\end{footnotesize}
\twocolumn
\begin{table}
\caption{Spectral lines observed in the Sr-filament of $\eta$~Car in the wavelength region 2480-10140~\AA. The lines are primarily sorted by ion and within each ion by excitation energy.}
\begin{tabular}{llllll} \hline \hline
Velocity	&	$\lambda_{lab}$ &	Ion &	Mult. &	Transition &	E$_{exc}$	\\
(km/s)	&	(\AA)	&		&		&		&	(eV) \\ \hline
-69	&	9826.822	&	[CI]	&	(1F)	&	2p$^2$~$^3$P$_1$-2p$^2$~$^1$D$_2$	&	1.3   \\
-83	&	9852.965	&	[CI]	&	(1F)	&	2p$^2$~$^3$P$_2$-2p$^2$~$^1$D$_2$	&	1.3   \\
-93	&	8729.523	&	[CI]	&	(3F)	&	2p$^2$~$^1$D$_2$-2p$^2$~$^1$S$_0$	&	2.7   \\
-115	&	5897.558	&	NaI	&	(1)	&	3s~$^2$S$_{1/2}$-3p~$^2$P$_{1/2}^o$	&	2.1   \\
-119	&	5891.583	&	NaI	&	(1)	&	3s~$^2$S$_{1/2}$-3p~$^2$P$_{3/2}^o$	&	2.1   \\
-95	&	2852.964	&	MgI	&	(uv1)	&	3s$^2$~$^1$S$_0$-3s3p~$^1$P$_1^o$	&	4.3   \\
\hline
\end{tabular}
\end{table}
\addtocounter{table}{-1}
\begin{table}
\caption{continued.}
\begin{tabular}{llllll} \hline \hline
Velocity	&	$\lambda_{lab}$ &	Ion &	Mult. &	Transition &	E$_{exc}$	\\
(km/s)	&	(\AA)	&		&		&		&	(eV) \\ \hline
-83	&	2541.585	&	ScII	&	(uv1)	&	a$^3$D$_1$-y$^3$P$_2$	&	4.9   \\
-105	&	2545.967	&	ScII	&	(uv1)	&	a$^3$D$_2$-y$^3$P$_2$	&	4.9   \\
-93	&	2553.120	&	ScII	&	(uv1)	&	a$^3$D$_3$-y$^3$P$_2$	&	4.9   \\
-90	&	8263.438	&	[ScII]	&	(3F)	&	a$^3$D$_1$-a$^3$P$_1$	&	1.5   \\
-85	&	8328.905	&	[ScII]	&	(3F)	&	a$^3$D$_2$-a$^3$P$_0$	&	1.5   \\
-96	&	8273.557	&	[ScII]	&	(3F)	&	a$^3$D$_2$-a$^3$P$_2$	&	1.5   \\
\hline
\end{tabular}
\end{table}
\addtocounter{table}{-1}
\begin{table}
\caption{continued.}
\begin{tabular}{llllll} \hline \hline
Velocity	&	$\lambda_{lab}$ &	Ion &	Mult. &	Transition &	E$_{exc}$	\\
(km/s)	&	(\AA)	&		&		&		&	(eV) \\ \hline
-101	&	3330.411	&	TiII	&	(7)	&	b$^4$F$_{7/2}$-z$^4$F$_{7/2}$	&	3.9   \\
-101	&	3309.757	&	TiII	&	(7)	&	b$^4$F$_{7/2}$-z$^4$F$_{9/2}$	&	3.9   \\
-80	&	3344.726	&	TiII	&	(7)	&	b$^4$F$_{9/2}$-z$^4$F$_{7/2}$	&	3.9   \\
-103	&	3323.897	&	TiII	&	(7)	&	b$^4$F$_{9/2}$-z$^4$F$_{9/2}$	&	3.9   \\
-115	&	5227.998	&	TiII	&	(70)	&	b$^2$D$_{3/2}$-z$^2$D$_{3/2}$	&	3.9   \\
-76	&	3162.136	&	TiII	&	(10)	&	b$^4$F$_{3/2}$-z$^4$D$_{1/2}$	&	4.0   \\
\hline
\end{tabular}
\end{table}
\addtocounter{table}{-1}
\begin{table}
\caption{continued.}
\begin{tabular}{llllll} \hline \hline
Velocity	&	$\lambda_{lab}$ &	Ion &	Mult. &	Transition &	E$_{exc}$	\\
(km/s)	&	(\AA)	&		&		&		&	(eV) \\ \hline
-79	&	3758.756	&	TiII	&	(73)	&	b$^2$D$_{3/2}$-z$^2$P$_{3/2}$	&	4.9   \\
-88	&	3777.124	&	TiII	&	(73)	&	b$^2$D$_{5/2}$-z$^2$P$_{3/2}$	&	4.9   \\
-80	&	4331.463	&	TiII	&	(93)	&	b$^2$P$_{1/2}$-y$^2$D$_{3/2}$	&	4.9   \\
-100	&	5212.987	&	TiII	&	(103)	&	b$^2$F$_{7/2}$-y$^2$F$_{7/2}$	&	5.0   \\
-105	&	3203.463	&	TiII	&	(26)	&	a$^2$D$_{3/2}$-y$^2$F$_{5/2}$	&	5.0   \\
-114	&	3191.802	&	TiII	&	(26)	&	a$^2$D$_{5/2}$-y$^2$F$_{7/2}$	&	5.0   \\
\hline
\end{tabular}
\end{table}
\addtocounter{table}{-1}
\begin{table}
\caption{continued.}
\begin{tabular}{llllll} \hline \hline
Velocity	&	$\lambda_{lab}$ &	Ion &	Mult. &	Transition &	E$_{exc}$	\\
(km/s)	&	(\AA)	&		&		&		&	(eV) \\ \hline
-116	&	3018.063	&	TiII	&	(85)	&	a$^2$H$_{11/2}$-z$^2$H$_{11/2}$	&	5.7   \\
-100	&	3009.188	&	TiII	&	(85)	&	a$^2$H$_{9/2}$-z$^2$H$_{11/2}$	&	5.7   \\
-94	&	3030.611	&	TiII	&	(85)	&	a$^2$H$_{9/2}$-z$^2$H$_{9/2}$	&	5.7   \\
-81	&	3288.608	&	TiII	&	(89)	&	b$^2$G$_{7/2}$-z$^2$H$_{9/2}$	&	5.7   \\
-105	&	3262.525	&	TiII	&	(89)	&	b$^2$G$_{9/2}$-z$^2$H$_{11/2}$	&	5.7   \\
-101	&	3090.306	&	TiII	&	(90)	&	b$^2$G$_{7/2}$-x$^2$F$_{5/2}$	&	5.9   \\
\hline
\end{tabular}
\end{table}
\addtocounter{table}{-1}
\begin{table}
\caption{continued.}
\begin{tabular}{llllll} \hline \hline
Velocity	&	$\lambda_{lab}$ &	Ion &	Mult. &	Transition &	E$_{exc}$	\\
(km/s)	&	(\AA)	&		&		&		&	(eV) \\ \hline
-102	&	3917.519	&	VII	&	(10)	&	a$^3$P$_1$-z$^3$D$_2$	&	4.6   \\
-94	&	3953.083	&	VII	&	(10)	&	a$^3$P$_2$-z$^3$D$_3$	&	4.6   \\
-95	&	2918.207	&	VII	&	(2)	&	a$^5$F$_2$-z$^5$F$_1$	&	4.6   \\
-104	&	2921.226	&	VII	&	(2)	&	a$^5$F$_2$-z$^5$F$_3$	&	4.6   \\
-107	&	2931.649	&	VII	&	(2)	&	a$^5$F$_3$-z$^5$F$_3$	&	4.6   \\
-94	&	2945.422	&	VII	&	(2)	&	a$^5$F$_4$-z$^5$F$_3$	&	4.6   \\
\hline
\end{tabular}
\end{table}
\addtocounter{table}{-1}
\clearpage
\begin{table}
\caption{continued.}
\begin{tabular}{llllll} \hline \hline
Velocity	&	$\lambda_{lab}$ &	Ion &	Mult. &	Transition &	E$_{exc}$	\\
(km/s)	&	(\AA)	&		&		&		&	(eV) \\ \hline
-91	&	2996.863	&	VII	&	(27)	&	a$^5$P$_1$-z$^5$P$_2$	&	5.8   \\
-90	&	3015.694	&	VII	&	(27)	&	a$^5$P$_2$-z$^5$P$_1$	&	5.8   \\
-90	&	3002.079	&	VII	&	(27)	&	a$^5$P$_3$-z$^5$P$_3$	&	5.8   \\
-91	&	3013.985	&	VII	&	(28)	&	a$^5$P$_1$-y$^5$D$_1$	&	5.8   \\
-81	&	3021.542	&	VII	&	(28)	&	a$^5$P$_2$-y$^5$D$_1$	&	5.8   \\
-79	&	2969.241	&	VII	&	(28)	&	a$^5$P$_3$-y$^5$D$_4$	&	5.9   \\
\hline
\end{tabular}
\end{table}
\addtocounter{table}{-1}
\begin{table}
\caption{continued.}
\begin{tabular}{llllll} \hline \hline
Velocity	&	$\lambda_{lab}$ &	Ion &	Mult. &	Transition &	E$_{exc}$	\\
(km/s)	&	(\AA)	&		&		&		&	(eV) \\ \hline
-99	&	2677.957	&	CrII	&	(uv8)	&	a$^6$D$_{9/2}$-z$^6$D$_{9/2}$	&	6.2   \\
-113	&	3121.264	&	CrII	&	(5)	&	a$^4$D$_{3/2}$-z$^4$F$_{5/2}$	&	6.4   \\
-90	&	3125.879	&	CrII	&	(5)	&	a$^4$D$_{5/2}$-z$^4$F$_{7/2}$	&	6.4   \\
-99	&	3132.961	&	CrII	&	(5)	&	a$^4$D$_{7/2}$-z$^4$F$_{9/2}$	&	6.4   \\
-117	&	8359.939	&	[CrII]	&	(1F)	&	a$^6$S$_{5/2}$-a$^6$D$_{1/2}$	&	1.5   \\
-129	&	8310.770	&	[CrII]	&	(1F)	&	a$^6$S$_{5/2}$-a$^6$D$_{3/2}$	&	1.5   \\
\hline
\end{tabular}
\end{table}
\addtocounter{table}{-1}
\begin{table}
\caption{continued.}
\begin{tabular}{llllll} \hline \hline
Velocity	&	$\lambda_{lab}$ &	Ion &	Mult. &	Transition &	E$_{exc}$	\\
(km/s)	&	(\AA)	&		&		&		&	(eV) \\ \hline
-105	&	3749.327	&	FeI	&	(5)	&	a$^5$D$_1$-z$^5$F$_2$	&	3.4   \\
-81	&	3746.626	&	FeI	&	(5)	&	a$^5$D$_2$-z$^5$F$_3$	&	3.4   \\
-87	&	3684.104	&	FeI	&	(5)	&	a$^5$D$_3$-z$^5$F$_2$	&	3.4   \\
-100	&	3738.194	&	FeI	&	(5)	&	a$^5$D$_3$-z$^5$F$_4$	&	3.4   \\
-101	&	3680.961	&	FeI	&	(5)	&	a$^5$D$_4$-z$^5$F$_4$	&	3.4   \\
-107	&	3441.592	&	FeI	&	(6)	&	a$^5$D$_4$-z$^5$P$_3$	&	3.6   \\
\hline
\end{tabular}
\end{table}
\addtocounter{table}{-1}
\begin{table}
\caption{continued.}
\begin{tabular}{llllll} \hline \hline
Velocity	&	$\lambda_{lab}$ &	Ion &	Mult. &	Transition &	E$_{exc}$	\\
(km/s)	&	(\AA)	&		&		&		&	(eV) \\ \hline
-87	&	2719.833	&	FeI	&	(uv5)	&	a$^5$D$_4$-y$^5$P$_3$	&	4.6   \\
-110	&	4072.887	&	FeI	&	(43)	&	a$^3$F$_2$-y$^3$F$_2$	&	4.7   \\
-70	&	4006.373	&	FeI	&	(43)	&	a$^3$F$_3$-y$^3$F$_2$	&	4.7   \\
-99	&	3816.923	&	FeI	&	(45)	&	a$^3$F$_4$-y$^3$D$_3$	&	4.7   \\
-94	&	3828.909	&	FeI	&	(45)	&	a$^3$F$_3$-y$^3$D$_2$	&	4.8   \\
-80	&	3058.335	&	FeI	&	(28)	&	a$^5$F$_5$-x$^5$D$_4$	&	4.9   \\
\hline
\end{tabular}
\end{table}
\addtocounter{table}{-1}
\begin{table}
\caption{continued.}
\begin{tabular}{llllll} \hline \hline
Velocity	&	$\lambda_{lab}$ &	Ion &	Mult. &	Transition &	E$_{exc}$	\\
(km/s)	&	(\AA)	&		&		&		&	(eV) \\ \hline
-101	&	5697.949	&	[FeI]	&	(2F)	&	a$^5$D$_4$-a$^5$P$_3$	&	2.2   \\
-91	&	5641.119	&	[FeI]	&	(2F)	&	a$^5$D$_4$-a$^5$P$_2$	&	2.3   \\
-105	&	4844.701	&	[FeI]	&	(4F)	&	a$^5$D$_4$-b$^3$F2$_4$	&	2.6   \\
-116	&	7007.180	&	[FeI]	&	(15F)	&	a$^5$F$_3$-a$^3$G$_4$	&	2.7   \\
-90	&	6838.838	&	[FeI]	&	(15F)	&	a$^5$F$_4$-a$^3$G$_4$	&	2.7   \\
-104	&	6974.000	&	[FeI]	&	(15F)	&	a$^5$F$_4$-a$^3$G$_5$	&	2.7   \\
\hline
\end{tabular}
\end{table}
\newpage
\begin{table}
\caption{Unidentified lines}
\begin{tabular}{ccl} \hline \hline
$\lambda_{obs}$ (\AA)	&	Int (nov01)$^a$ 	& Wavelength coincidence \\ \hline
2537.92	&	14	& bl	\\		
2538.90	&	22	&	\\		
2548.76	&	7	& VII a$^3$P$_1$-y$^3$P$_1$ \\		
2580.24	&	37	&	\\		
2592.96	&	33	& bl abs	\\		
2815.20	&	3$^b$	& FeI a$^3$F$_2$-sp$^5$H$_3$ \\		
3050.49	&	8	& VII  a$^3$D$_3$-y$^3$P$_2$ \\		
3146.64	&	5	& 	\\		
3208.63	&	8	&	\\		
3379.87	&	19	& TiII b$^4$P$_{5/2}$-y$^2$D$_{3/2}$ 	\\
3391.76	&	15	& ZrII a$^4$F$_{9/2}$-z$^4$G$_{11/2}$	\\		
3414.65	&	7	& VII b$^3$D$_3$-z$^1$D$_2$	\\		
3416.15	&	13	&	\\		
3438.51	&	34	& ZrII a$^4$F$_{7/2}$-z$^4$G$_{9/2}$, bl. \\		
3600.62	&	5	& YII a$^3$D$_3$-z$^3$D$_3$	\\		
3687.65	&	11	&	\\		
3697.79	&	4	&	\\		
3709.35 &       9       & YII a$^3$D$_3$-z$^3$F$_3$, bl. \\
3774.24	&	9	& TiII a$^2$F$_{5/2}$-z$^4$F$_{7/2}$ \\		
	&		& YII a$^3$D$_2$-z$^3$F$_3$	\\		
	&		& SrII 5d $^2$D$_{5/2}$-7f $^2$F	\\		
3856.35	&	4	& 	\\		
3958.07 &       1       & ZrII a$^2$D$_{3/2}$-z$^2$F$_{5/2}$ \\ 
4305.24	&	6	& SrII 5p $^2$P$_{3/2}$-6s $^2$S$_{1/2}$	\\		
4427.25	&	2	& TiII b$^4$P$_{5/2}$-z$^4$D$_{3/2}$ \\		
4747.17	&	9	& SrII 6s $^2$S$_{1/2}$-7p $^2$P$_{1/2}$ \\		
4763.86	&	4	& TiII a$^2$P$_{3/2}$-z$^4$F$_{5/2}$ \\		
4802.32	&	7	& 	\\		
5253.20	&	1	& FeI z$^5$D$_1$-e$^5$D$_1$	\\		
5303.74	&	3	& [FeI] a$^5$D$_2$-a$^3$P$_1$	\\		
5470.63	&	7	& 	\\		
5559.85	&	4	& \\		
5599.35	&	1	& 	\\		
6160.44	&	2	& \\		
6219.08	&	2	& FeI a$^5$P$_2$-y$^5$D$_2$ \\		
6998.68	&	2	&	\\		
7426.27	&	7	&	\\		
7430.71	&	4	&	\\
7664.71	& 	13	& YII a$^3$D-a$^3$P \\	
7749.18	&	4	&	\\		
7808.06	&	6	& \\		
8050.43	&	3	&	\\		
8674.12	&	2	& [VII] a$^3$F$_4$-b$^3$D$_3$\\		
8698.34	&	2	& [VII] a$^5$F$_5$-a$^3$G$_5$	\\		
8806.31	&	4	&	\\		
9341.82	&	4	& \\ \hline 		
\end{tabular}

$^a$ 10$^{-15}$ erg cm$^{-2}$ s$^{-1}$ arcsec$^{-2}$\\
$^b$ The 2815 \AA\ line is not covered in the Nov01 observations. The observed intensity is from the Mar00 data. 

\end{table}

\end{document}